\documentclass{PoS}
\pdfoutput=1

\usepackage{wg5-macros}

\title{Working Group 5: Physics with Heavy Flavours}

\ShortTitle{WG5: Physics with Heavy Flavours}

\author{%
Andrea Giammanco\\
Centre for Cosmology, Particle Physics and Phenomenology, Universit\'{e} catholique de Louvain, Belgium\\
E-mail: \email{andrea.giammanco@uclouvain.be}
}

\author{%
Rhorry Gauld\\
Institute of Theoretical Physics, ETH Z\"urich, Switzerland\\
E-mail: \email{rgauld@phys.ethz.ch}
}

\author{%
\speaker{Alex Pearce}\\
CERN, Switzerland\\
E-mail: \email{alex.pearce@cern.ch}
}

\abstract{%
This paper summarises a few selected topics discussed during Working Group 5 of DIS'17, Physics with Heavy Flavours, related to the study of charm, bottom, and top quark physics. 
While the programme of this Working Group was structured by thematic areas, this conference was the occasion for intense cross-pollination between traditionally disjoint research lines.
The four LHC experiments all contribute to heavy-flavour physics, with some degree of overlap in most areas, while experiments at other accelerators provide vital input in complimentary kinematic regions.
Theorists now have the possibility to take inputs from more sources, and experimentalists focus on measurements that maximise utility.
The interplay of LHC heavy quark cross-section measurements with DIS expertise is greatly improving PDF precision, leading to much improved models that, amongst other things, better inform the prospects for future colliders.
}

\FullConference{%
XXV International Workshop on Deep-Inelastic Scattering and Related Subjects\\
3--7 April 2017\\
University of Birmingham, UK
}

\begin{document}

\section{Introduction}

The Working Group 5 of DIS'17, Physics with Heavy Flavours, featured 41 presentations, six of them in common with Working Group 1, Structure Functions and Parton Densities.
Several experimental and theoretical aspects related to the production and properties of the so-called `heavy quarks', namely charm, bottom, and top, were discussed during eight sub-sessions devoted to the following topics: top quark pair production;
    single top quark production and properties;
    top quark mass;
    beauty and charm quark production;
    heavy flavour measurements as inputs for PDF fits (joint with Working Group 1);
    properties of \B and \D hadrons;
    exotic states; and
    quarkonia.
Given the required brevity of these proceedings, we cannot do justice to the broad scope of this programme that our speakers covered so expertly. We instead briefly touch on a few selected topics that span this list, and strongly encourage the reader to find more details in the corresponding individual contributions to the DIS'17 proceedings.
 
Section~\ref{sec:production} is devoted to the measurements of heavy-quark cross sections, inclusive or differential and for different colliding particles, while Section~\ref{sec:input} shows examples of usage of these measurements as inputs for the extraction of model parameters, such as in global proton parton density function~(PDF) fits and heavy quark mass determinations.
In Section~\ref{sec:properties}, related to heavy quark properties, we chose to highlight recent measurements of heavy-hadron decay rates and some anomalies in the \bottomq-quark sector.
Section~\ref{sec:exotica} addresses the current situation in the study of exotic states, i.e.\ bound states of 4 or more quarks.
Section~\ref{sec:quarkonia} discusses the production measurements of quarkonia, with particular focus on the effects of double parton scattering.

\section{Heavy quark production}
\label{sec:production}

\subsection{Top quark production}

The top quark is a particularly unique object: it is the heaviest known elementary particle. As a consequence of its large mass, it can decay weakly via an on-shell $W$ boson on extremely short time scales before the effects of decoherence or hadronisation can occur.
On the order of $10^{7}$ \ttbar\ pairs have been produced in \pp collisions at the LHC so far. Both inclusive and differential measurements of \ttbar production have reached a remarkable precision for all decay channels at centre-of-mass energies of 7, 8, and $13\,$TeV, and these new data, coupled to steadfast advances in theory accuracy~\cite{scott_dis17}, have led to an improved understanding of proton structure (including the photon PDF~\cite{tsinikos_dis17}) as well as stronger constraints on Standard Model parameters such as \alphas and \mtop, as elaborated in Sec.~\ref{sec:input}.
  
\begin{figure}[!h!tbp]
  \includegraphics[width=0.45\textwidth]{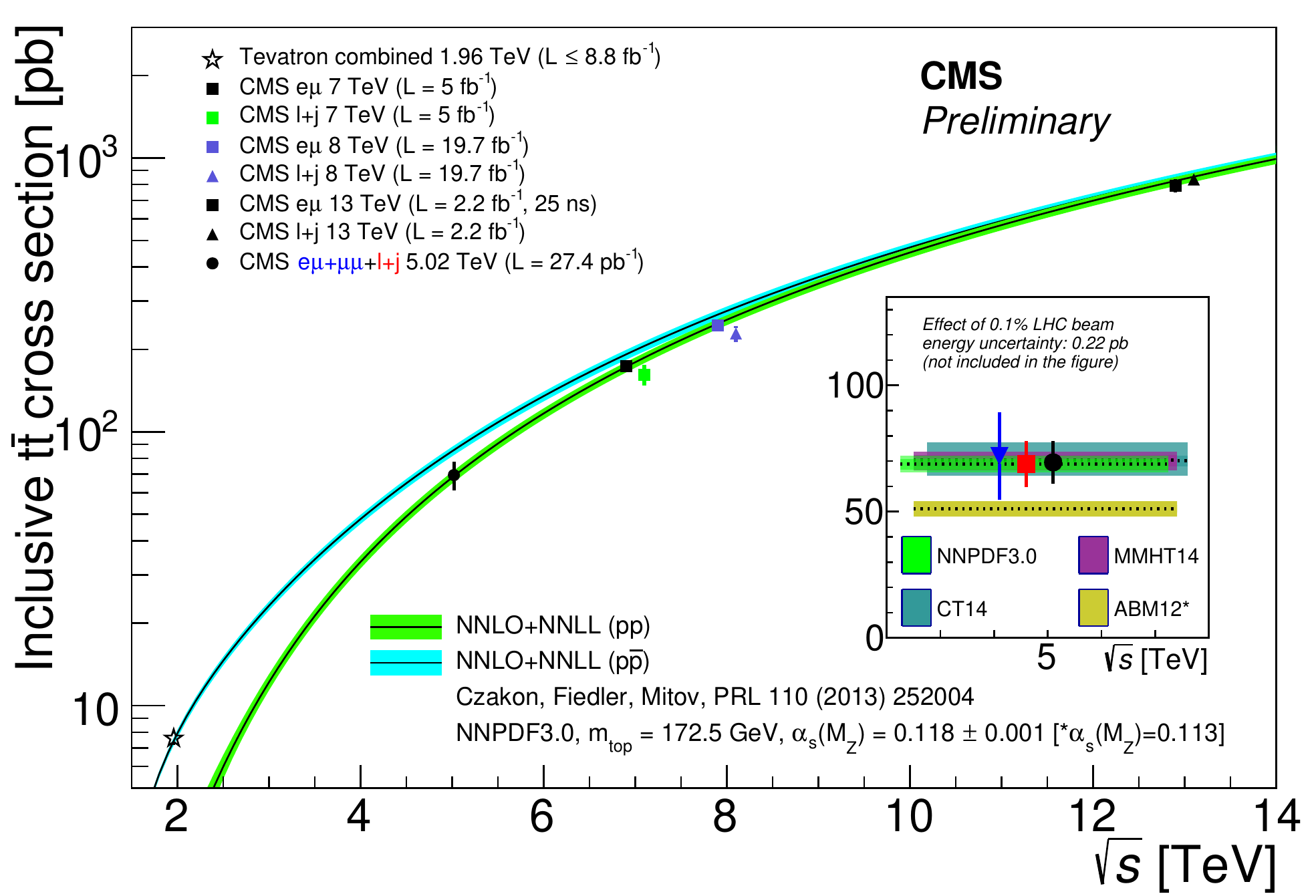}
  \hfill
  \includegraphics[width=0.45\textwidth]{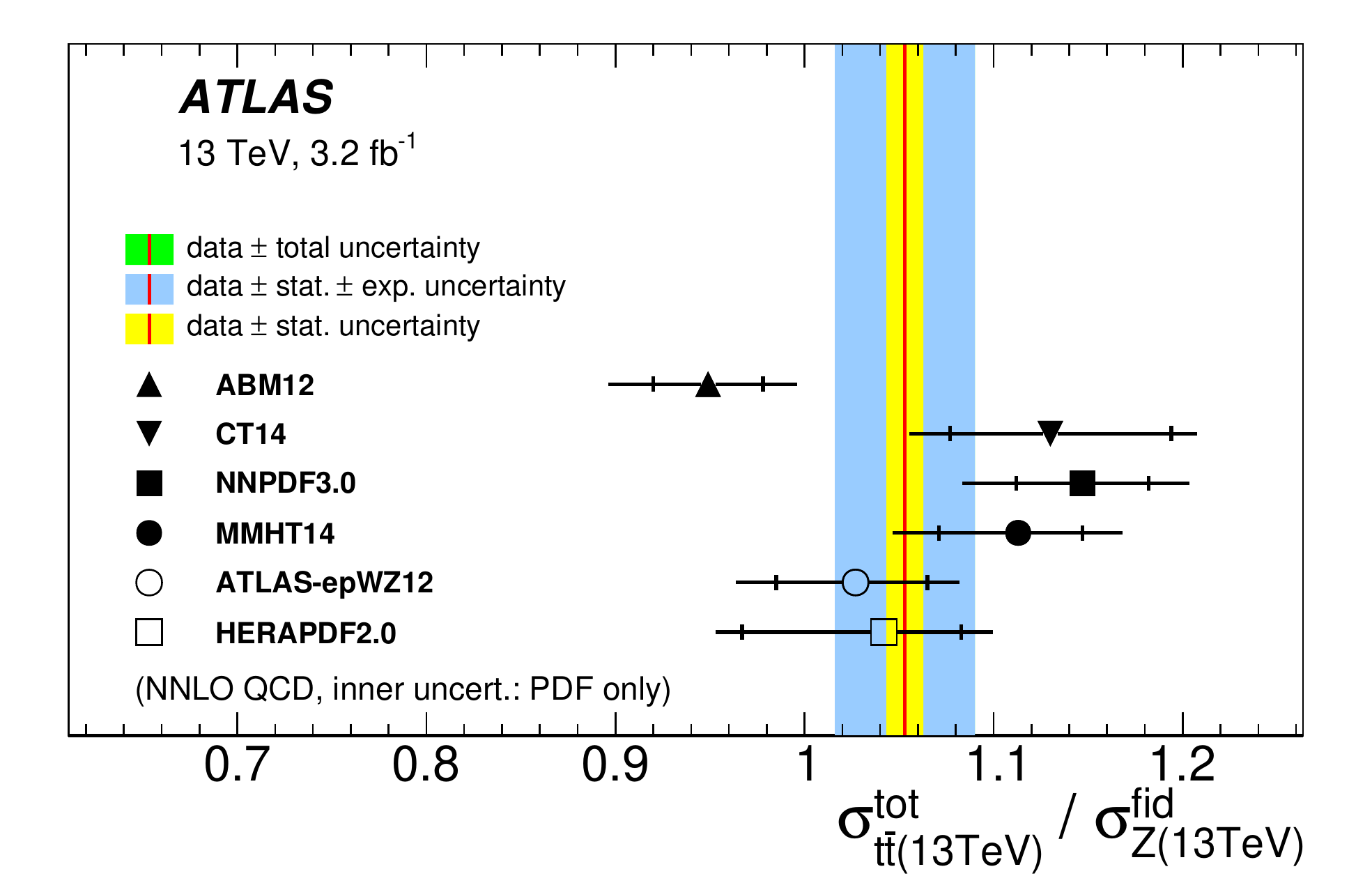}
  \caption{(Left) Summary of the \ttbar cross section measurements by the CMS collaboration. (Right) Measurement of $\sigma(\ttbar)/\sigma(Z)$ at \sqrtseq{13} by the ATLAS collaboration~\cite{Aaboud:2016zpd}.}
  \label{fig:ttbar-xs}
\end{figure}

A new result released at the time of this conference~\cite{gonzalez_dis17} has been the first measurement of the inclusive \ttbar cross section at \sqrtseq{5.02}~\cite{CMS-PAS-TOP-16-023} by CMS, see Figure~\ref{fig:ttbar-xs} (left), exploiting the data from a brief \pp run with an integrated luminosity of $27.4\,$pb$^{-1}$. This measurement combines data samples with either one or two reconstructed leptons and achieves a relative precision of 12\%, dominated by the statistical uncertainty. The impact of this measurements on large-$x$ PDFs is discussed in Sec.~\ref{sec:pdfs}.

Another interesting result is the measurement of top quark pairs normalised to that of $Z$ boson production, which was presented by the ATLAS collaboration~\cite{yamazaki_dis17,Aaboud:2016zpd} at 7, 8, and $13\,$TeV.
The motivation is that as large data samples are obtained at the LHC, the corresponding measurements of absolute cross-sections will become limited by knowledge of the luminosity. As both $Z$-boson cross section measurements and the corresponding theoretical predictions are extremely precise, this channel can be used in ratio with other processes to eliminate the uncertainties due to the luminosity, which are fully correlated.
The current data already demonstrate significant power to constrain both the gluon as well as the the light-quark sea PDFs for $x$-values of $~0.1$ as well as $x < 0.02$ respectively. The measured cross section ratio at $13\,$TeV is compared to NNLO QCD predictions obtained with state-of-the-art PDF sets in Fig.~\ref{fig:ttbar-xs} (right).

\subsection{Beauty and charm production}

The production of charm and beauty quarks at the LHC occurs at very large rates, and the experimental
collaborations have been extremely successful in performing precise measurements of the production
cross-sections of these heavy quarks in \pp, \pPb, and \PbPb collisions. 
These various measurements are important for understanding the internal structure of the colliding hadrons, as well as testing for the formation of a quark-gluon-plasma~(QGP) within the dense environment of \PbPb collisions. The remainder of this Section will focus on the results relevant for \pPb and \PbPb collisions. 
The impact of the measurement of heavy quark production in \pp collisions on free nucleon PDFs will be discussed in
the following Section.

\begin{figure}[!h!tbp]
  \includegraphics[width=0.45\textwidth]{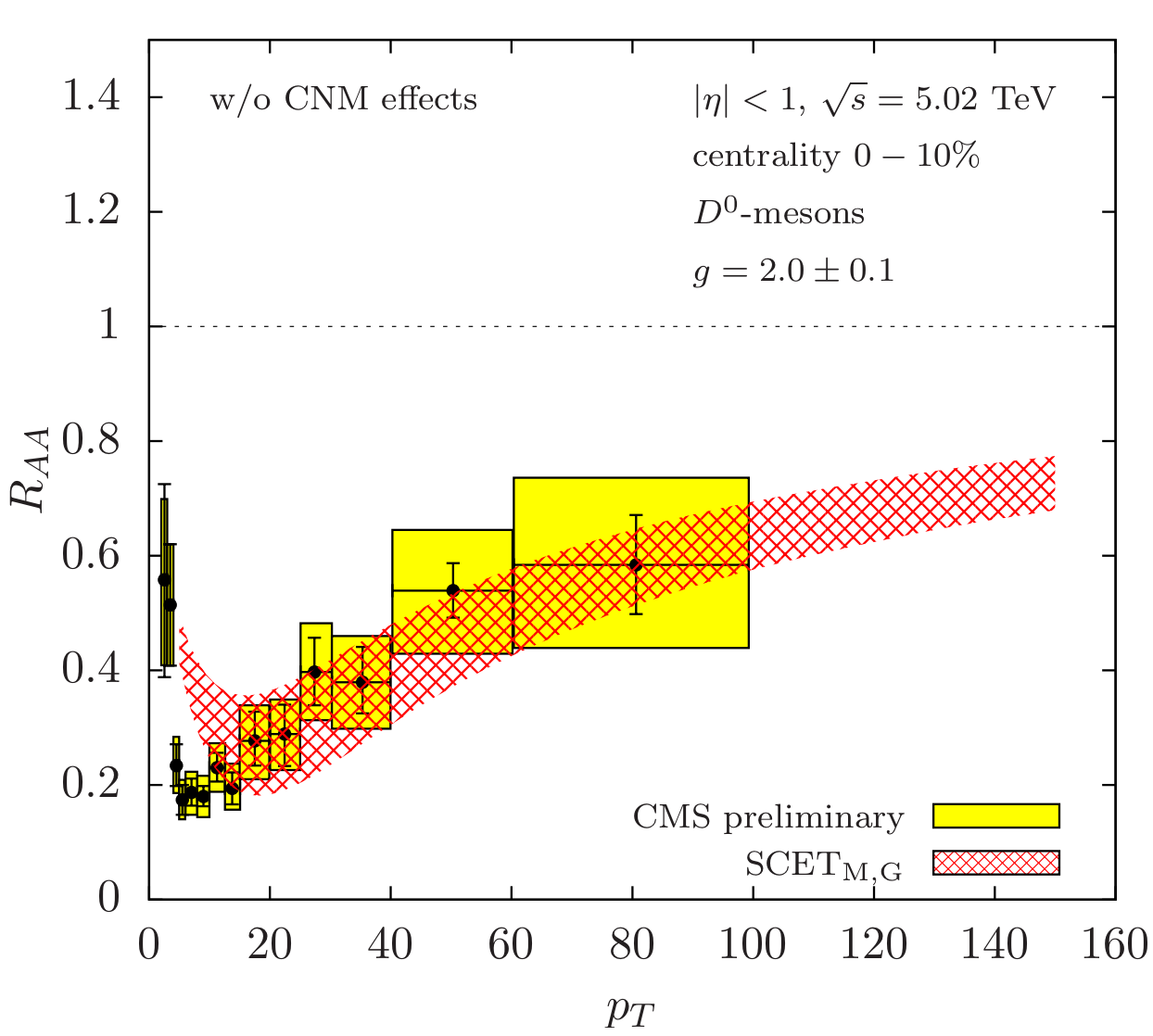}
  \hfill
  \includegraphics[width=0.45\textwidth]{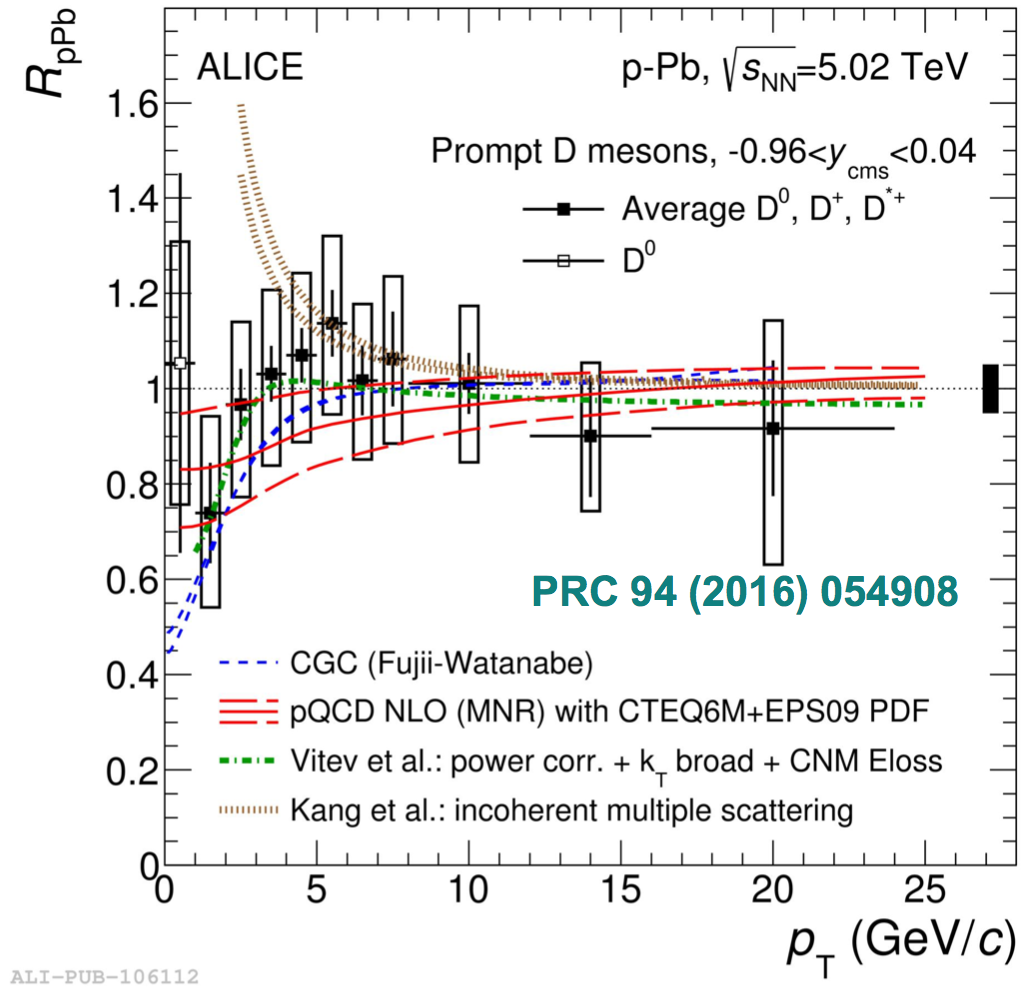}
  \caption{(Left) Supression of \D hadron production observed in \PbPb collisions by CMS, as compared to SCET predictions~\cite{Kang:2016ofv}. (Right) Measurement of CNM effects in \D production by ALICE in \pPb collisions~\cite{Adam:2016ich}.}
  \label{fig:heavy-ions}
\end{figure}

Heavy quarks are produced on relatively short time-scales in the hadronic scattering process. If a nuclear medium is formed in the process, the heavy quarks therefore have the opportunity to interact with such a medium.
A distinct signal of this type of interaction (which is expected in \PbPb collisions) is the suppression of heavy quark production with respect to a reference cross section measurement in \pp collisions.
Recent theoretical progress in the modelling of these effects has been made in Ref.~\cite{Kang:2016ofv}, where the interaction of heavy quarks with a nuclear medium has been modelled in the framework of soft collinear effective field theory (SCET). A comparison of these predictions for \D hadrons with the corresponding CMS data is shown in Fig.~\ref{fig:heavy-ions} (left). 
In addition to theoretical progress in modelling these intra-medium effects, it is also necessary to better understand cold nuclear matter~(CNM) effects which effectively describe the difference of collisions of bound nucleons with respect to free nucleons.
The study of the formation of a QGP in \PbPb collisions necessarily requires an understanding of these CNM effects, which can lead to a similar suppression of the rate of heavy quark production.
The impact of the CNM effects on \D hadron production has been studied by the ALICE collaboration~\cite{Adam:2016ich}, where the cross section observed in \pPb collisions is shown normalised to that in free \pp collisions in Fig.~\ref{fig:heavy-ions}. The available data are consistent with the size of CNM effects predicted by several models, and more precision is required to differentiate between these models.  

\section{Heavy quarks as inputs}
\label{sec:input}

The precise experimental measurements of heavy quark production at the LHC,
as well as those performed at HERA, can be used to extract a range of important information.
In this Section we first focus on how these measurements can be used to extract
information on proton structure, before discussing the determination of both
the top and charm quark masses.

\subsection{Heavy quarks and PDFs}
\label{sec:pdfs}

\begin{figure}[!h!tbp]
  \includegraphics[width=0.45\textwidth]{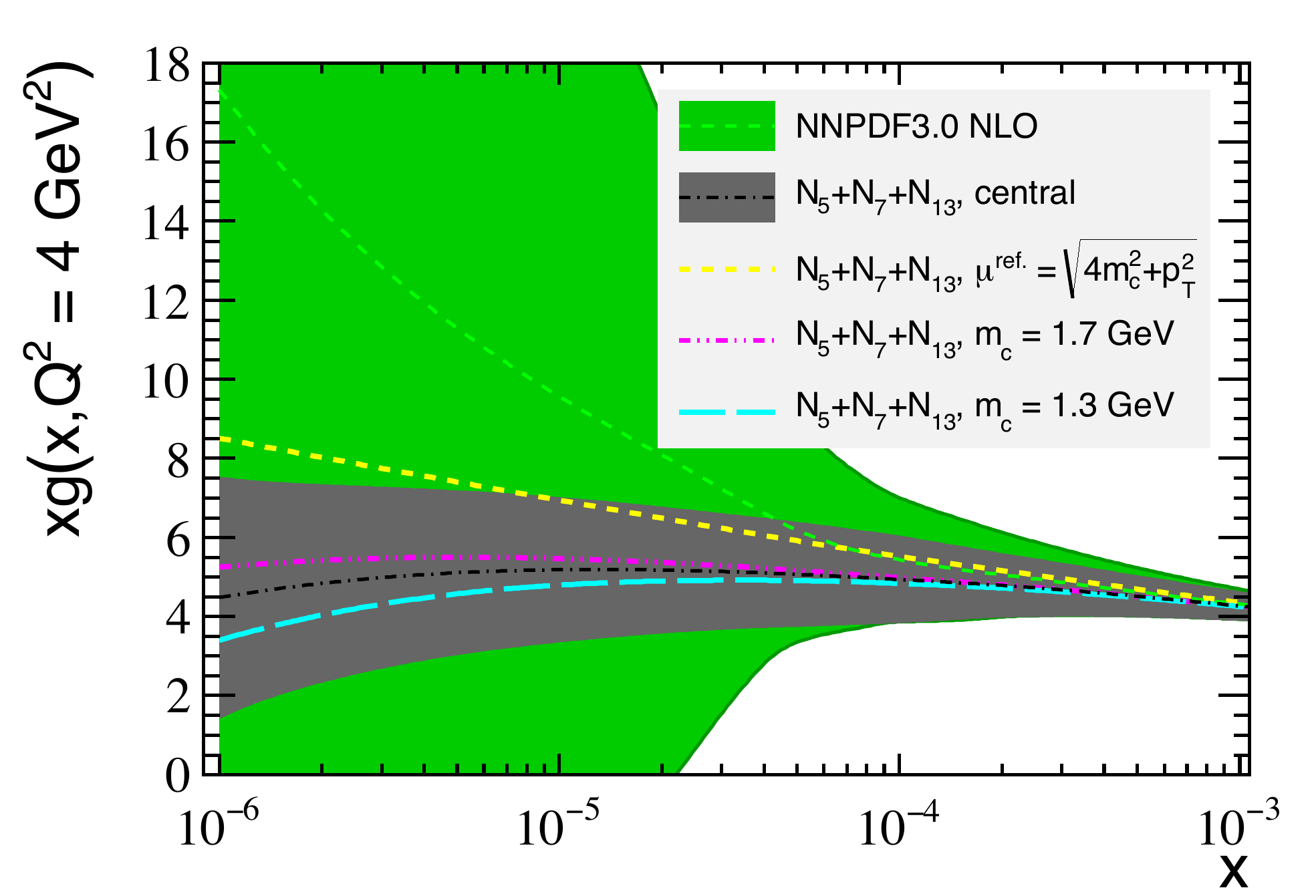}
  \hfill
  \includegraphics[width=0.45\textwidth]{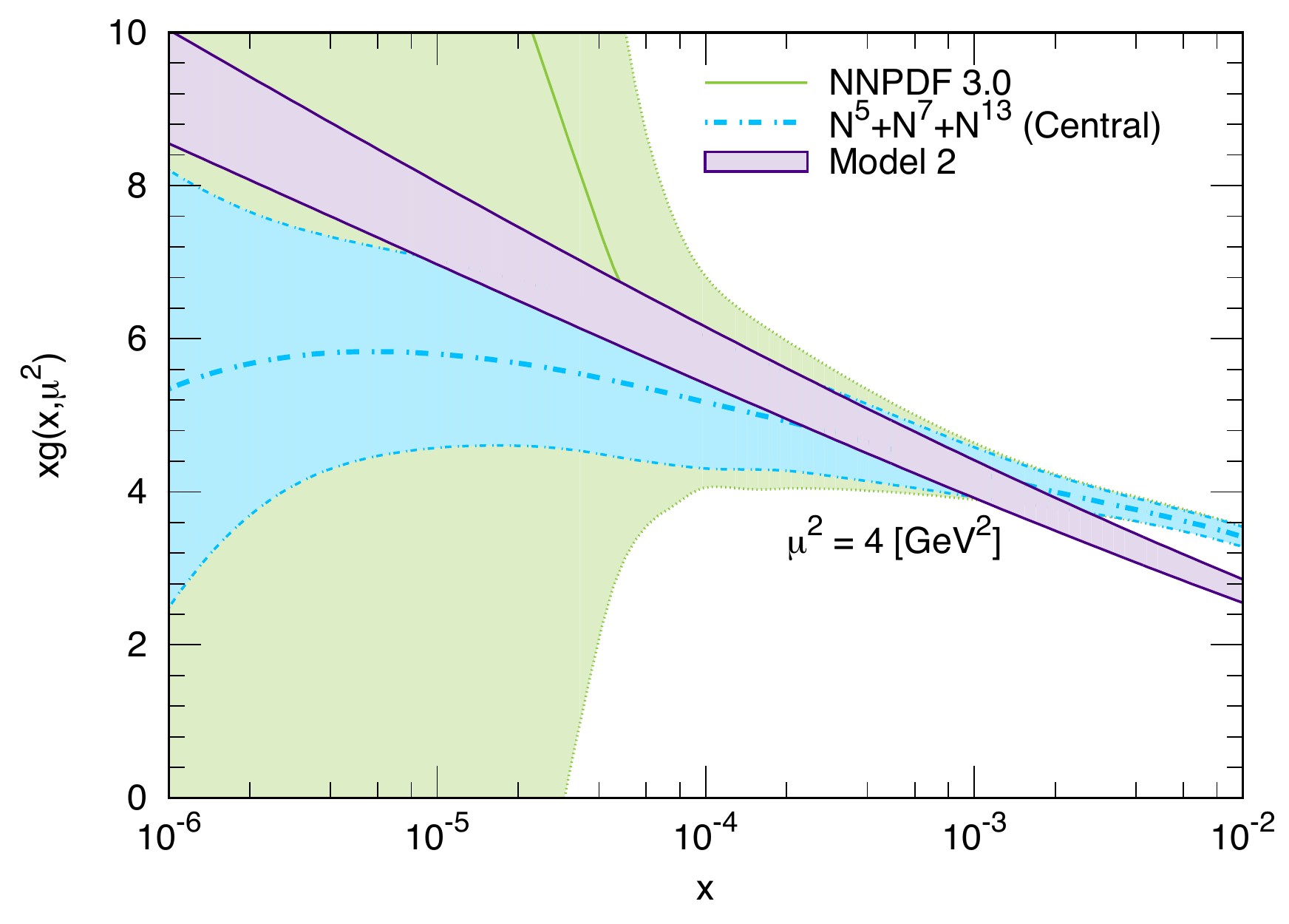}
  \caption{(Left) The gluon PDF obtained from including LHCb charm production data
  into the NNPDF3.0 global analyses~\cite{Gauld:2016kpd}. (Right) The gluon PDF obtained from an analysis of exclusive $J/\psi$ production~\cite{Jones:2016icr}.}
  \label{fig:pdf-low-x}
\end{figure}

The production of charm and beauty quarks in $ep$-collisions at HERA provides a direct probe of the gluon PDF,
and the corresponding measurements of this process have for some time now been an important data set for understanding the behaviour of the gluon for $x$-values as low as $3\cdot10^{-5}$ (for $Q^2 > 2{\rm~GeV}^2$)~\cite{Abramowicz:1900rp}.
At the LHC, the production of heavy quark pairs also provides a probe of the gluon PDF, however
covering much more extreme kinematics ranges in both $x$ and $Q^2$. This kinematic dependence, at leading order, is given by
\begin{equation}
    x_{1,(2)} = \frac{m_T^Q}{\sqrt{s}} \left( e^{(-) y_Q} + e^{(-) y_{\bar{Q}}} \right)\,,
\end{equation}
where $m_T^Q$ is the transverse heavy quark mass, $\sqrt{s}$ is the hadronic centre-of-mass energy, and $y_{Q,\bar{Q}}$ are the outgoing heavy quark rapidities.
It is therefore possible to access extremely large (small) values of $x$ by considering top (charm) quark production at the LHC, which in turn provides useful information for a range of different physics process both at the LHC as well as in the field of neutrino astronomy.
A recent analysis of LHCb cross section measurements of $D$ hadron production at various centre-of-mass energies
was recently performed in Refs.~\cite{rojo_dis17,Gauld:2016kpd}. This data extends the reach of global analyses of collinear PDFs, and
leads to a substantial reduction in the uncertainties in a region of $x < 10^{-5}$ as shown in Fig.~\ref{fig:pdf-low-x} (left)
at $Q = 2$~GeV. In addition, an analysis of exclusive \Jpsi production has also been performed in Refs.~\cite{jones_dis17,Jones:2016icr}. 
A direct comparison of these results is not possible as the PDFs are extracted within different theoretical frameworks, however an approximate transformation of the latter results allows for a comparison at the level of collinear PDFs and is shown in Fig.~\ref{fig:pdf-low-x} (right).
Improvements in the understanding of the low-$x$ gluon PDF have important consequences for a
number physics processes such as the ultra-high-energy neutrino-nucleon cross section or the production 
of charm quarks within Earth's upper atmosphere, the latter of which can lead to a significant background in the measurements
of high-energy astrophysical neutrino at neutrino telescopes.

While the LHC charm data leads to improvements in the understanding of the gluon PDF at low-$x$, the wealth
of precise top quark production data can also be used to constrain the gluon PDF large-$x$.
An analysis of how the singly differential \ttbar cross section measurements at 8~TeV by ATLAS and CMS
impact the gluon PDF constraints was performed at NNLO~\cite{nocera_dis17,Czakon:2016olj}. The impact of the inclusive 
\ttbar cross section measurement at 5.02~TeV has also been studied by CMS~\cite{gonzalez_dis17,CMS-PAS-TOP-16-023}. These data lead to important constraints on the behaviour of the gluon PDF at large-$x$ which has consequences for
improving the sensitivity of new physics searches at large partonic centre-of-mass energies. 
The CMS collaboration has also recently presented the first double differential \ttbar cross-section
measurements, using 8~TeV data, which also lead to similar constraints on the large-x gluon PDF~\cite{gonzalez_dis17,Sirunyan:2017azo}.

\subsection{Heavy quark mass determinations from cross sections}

As the predicted rate of heavy quark production (both in $pp$- and $ep$-collisions) depends on the input value of the heavy quark mass, heavy quark cross measurements can be used to directly constrain the heavy quark masses.

Precise determination of the value of these heavy quark masses is important for performing tests of the Standard Model, an important example being the inter-consistency of parameters describing the top quark, $W$ boson and Higgs boson masses.
Such a test is performed in a well-defined renormalisation scheme, and requires that the input values from experiment (top quark mass, etc.) have also been extracted in a similar manner.

A topic which has received much theoretical attention~\cite{corcella_dis17} in this regard, is the translation or interpretation of the top quark mass, which is experimentally extracted using standard Monte Carlo Parton Shower predictions --- the mass extracted in this way is often referred to as the ``MC mass''.
  
Traditional mass measurements~\cite{kovalchuk_dis17,melini_dis17} aim at reconstructing the decay products of the top quark, while several recent results have been produced with alternative approaches that minimise the dependence on non-perturbative effects. While these new methods are not yet competitive in precision with the standard ones, they have complementary systematic uncertainties and serve as successful proofs of principle that pave the way to future measurements with larger statistics.

In particular, the top-quark pole mass can be extracted from \ttbar cross section measurements, either inclusive or differential~\cite{heymes_dis17,d0-note-6473-CONF}.
Another alternative is to simultaneously fit for $\sigma(\ttbar)$ and $m_{t}^{\rm{MC}}$~\cite{kieseler_dis17,Kieseler:2015jzh}, using a distribution whose shape is directly sensitive to the MC mass while the normalisation, being directly proportional to the cross section, gives sensitivity to the pole mass. This method effectively provides a ``calibration'' of the MC mass to the pole mass.

In addition to recent progress in top-quark mass determinations, new results 
were also shown for charm-quark mass extractions. In a recent 
analysis~\cite{geiser_dis17,Gizhko:2017fiu}, the rate of charm quark production 
in $ep$ collisions has been used to extract the \msbar~scheme charm-quark mass 
across a wide range in $Q^2$. These data provide complimentary (and consistent) 
results with respect to previous low energy extractions of the charm-quark 
mass.

\section{Properties of heavy flavour hadrons}
\label{sec:properties}

Hadrons containing heavy quarks exhibit a rich phenomenology, from neutral \D and \B meson mixing to the violation of the \CP symmetry via the Cabibbo-Kobayashi-Maskawa~(CKM) quark mixing mechanism. In addition, the decay rates and amplitudes of suppressed processes, of which there are many, are sensitive to contributions from physics beyond the Standard Model, offering probes that can reach beyond the scales of the hard processes. Heavy flavour measurements at frontier colliders are then complementary to direct searches.

In recent history, a majority of the most precise measurements and tests of charm and beauty hadron properties had come from the \belle and \babar experiments, each operating at \epem colliders running at a centre-of-mass energy close to the \UfourS mass, just above the $\B\Bbar$ threshold. This provides a very clean laboratory for study, at the cost of low production cross-sections. This is in contrast with the experiments at the LHC, where both the \ccbar and \bbbar production cross-sections are orders of mangitude larger, albeit accompanied by larger backgrounds. The LHCb experiment is the dedicated heavy flavour experiment at the LHC, although ALICE, ATLAS, and CMS also make valuable contributions to the field. Together, they are often measuring uncertainties below those of \belle and \babar. \besthree represents the current generation of \epem collider experiments, operating within a spectra of centre-of-mass energies around the di-charm thresholds ($\D\D$ and $\Lc\Lc$). \belletwo is expected to begin data-taking in 2018.

\begin{figure}[!h!tbp]
  \begin{center}
    \includegraphics[width=0.5\textwidth]{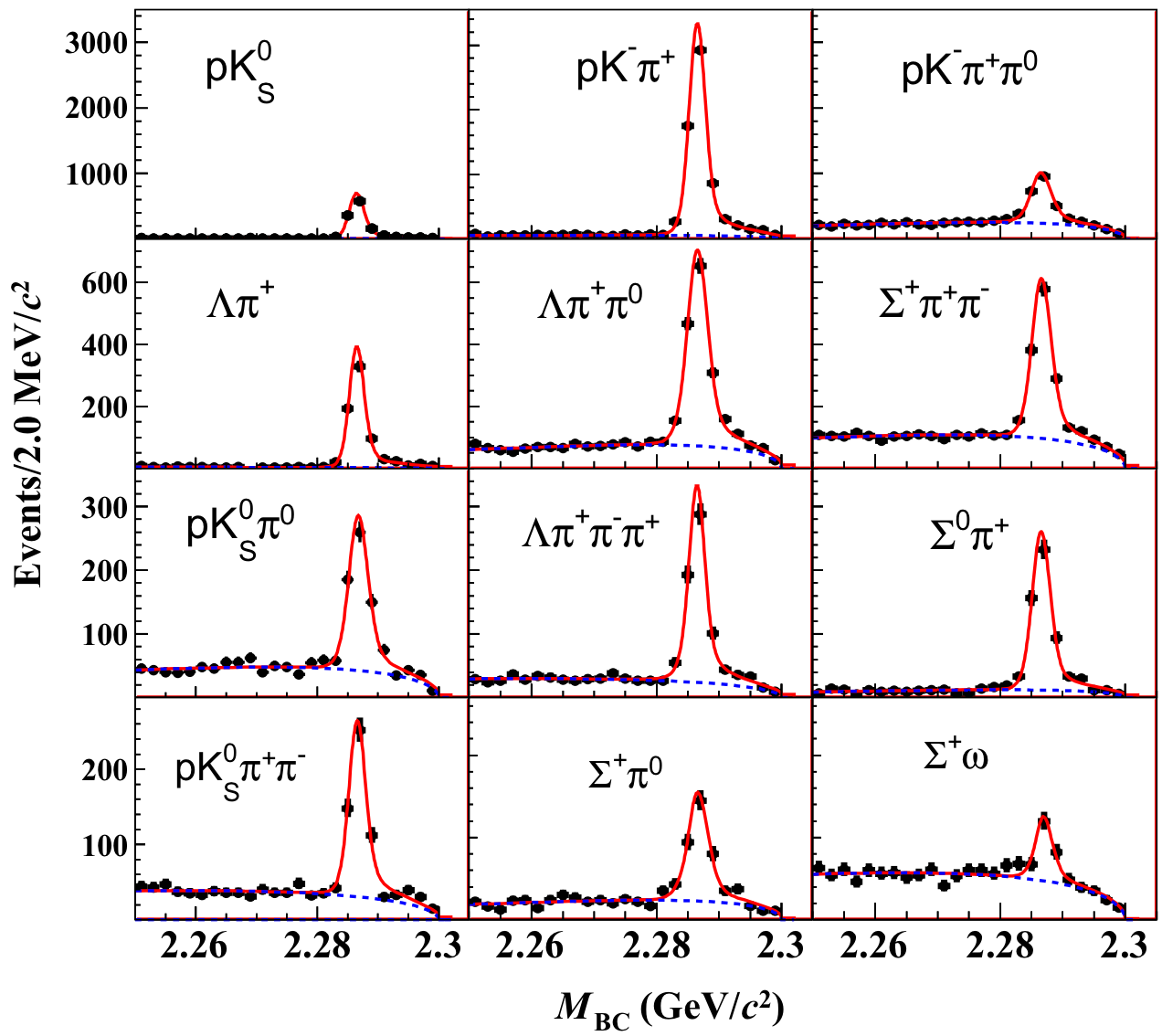}
  \end{center}
  \caption{%
    Mass distributions for twelve final states, made as part of a suite of measurements of absolute \Lcp branching fractions by the \besthree collaboration~\cite{Ablikim:2015flg}.
  }
  \label{fig:lc}
\end{figure}

The properties of charm baryons can play an important role in some studies of beauty decays. A recent measurement of the CKM matrix element $|\Vub|$ by LHCb~\cite{Aaij:2015bfa} used the semileptonic $\Lbz \to  p\mum\antinumu$ decay as a signal, and the $\Lcp\mum\antinumu$ final state as a control channel with $\Lcp \to p\Km\pip$. At that time, the branching fraction of this $\Lcp$ decay carried a $25\,\%$ relative uncertainty, which carried through as the leading systematic on the $|\Vub|$ measurement. The \besthree experiment recently published measurements of twelve absolute \Lcp branching fractions for Cabibbo-favoured fully hadronic final states, shown in Fig.~\ref{fig:lc}, as well as those for Cabibbo-suppressed modes and semileptonic decays~\cite{Ablikim:2015flg}. This impressive set includes the $p\Km\pip$ decay, for which a relative uncertainty of $6\,\%$ was reported. The precision for many of the other modes was also a significant improvement on the world averages.

\begin{figure}[!h!tbp]
  \begin{center}
    \includegraphics[width=0.5\textwidth]{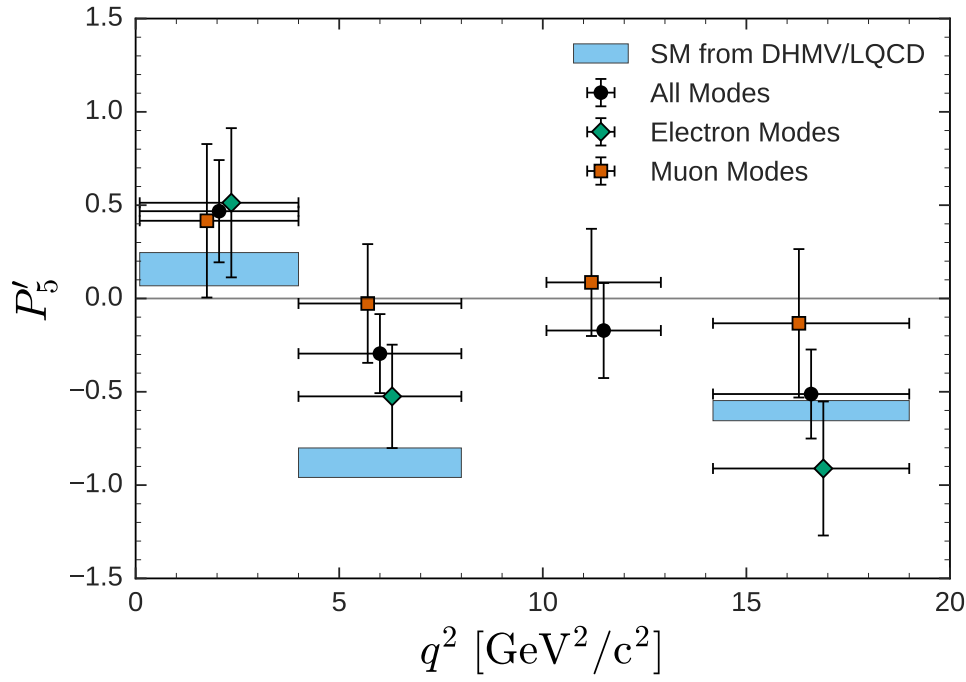}
  \end{center}
  \caption{%
    Distributions of the $P_{5}'$ observable as measured by the \belle experiment (points) and as predicted by a theoretical model that only includes SM effects~\cite{Wehle:2016yoi}. The second bin corresponds to the same region in which the LHCb experiment found a significant local deviation from theory.
  }
  \label{fig:p5prime}
\end{figure}

Arguably the hottest topic in beauty decays as of DIS'17 is the family of flavour-changing neutral current~(FCNC) processes $b \to s\ellp\ellm$. In the SM these can only proceed via Feynman diagrams including loops, and so their amplitudes contain a loop suppression factor as well as a factor of $|V_{ts}|$, such that their decay rates are particularly sensitive to BSM effects which are suppressed at high mass scales. In 2015, the LHCb experiment confirmed a deviation from the SM expectation of over $3\,\sigma$ in the angular distribution of the $\Bz \to \Kstz\mup\mum$ decay using their full Run 1 dataset~\cite{Aaij:2015oid}. The CMS experiment presented their own results at DIS'17~\cite{CMS:2017ivg}, using the \sqrtseq{8} dataset, which are in agreement with the SM. \belle also presented results on $\Kstz\ellp\ellm$~\cite{Wehle:2016yoi}, uniquely providing information on both the dimuon and dielectron modes, allowing not only for study of the individual angular distributions but also of their relative shapes, probing possibles effects from lepton flavour non-universality~(LFU). \belle finds a $2.6\,\sigma$ discrepancy with the SM using the dimuon mode, in the same region as LHCb, but only a $1.3\,\sigma$ difference for the dielectron, as shown in Fig.~\ref{fig:p5prime}. The tests of LFU are compatible with lepton flavour symmetry.

\begin{figure}[!h!tbp]
  \begin{center}
    \includegraphics[width=0.5\textwidth]{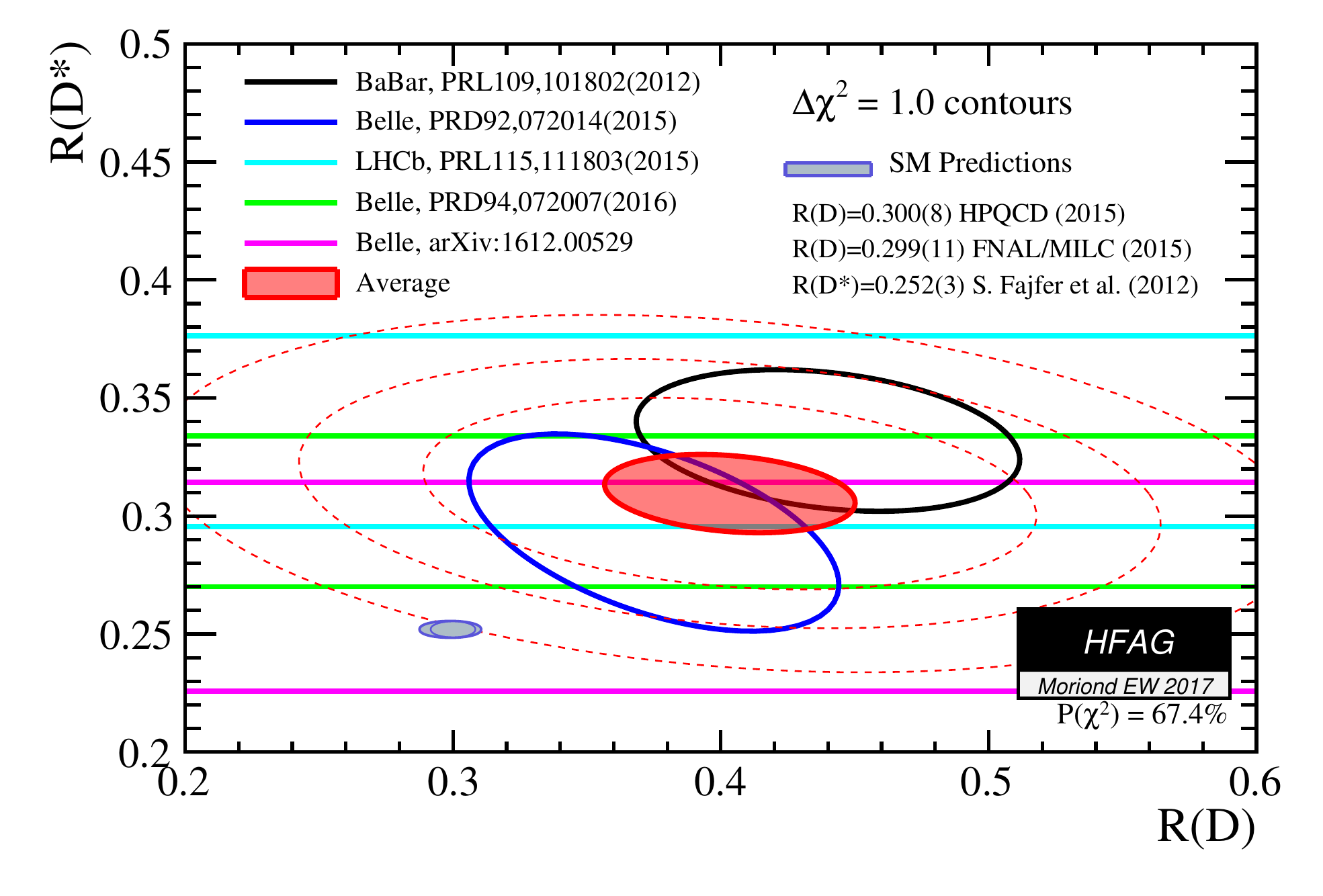}
  \end{center}
  \caption{%
    The current world average (red filled ellipse) of $R(D)$ and $R(D*)$ in the 2D plane alongside the various experimental inputs (hollow bands and ellipses) in contrast to SM predictions (blue filled ellipses)~\cite{Amhis:2016xyh}.
  }
  \label{fig:rdrds}
\end{figure}

Although tantalising, the discrepancies between experimental data and theoretical predictions in $b \to s\ellp\ellm$ processes are unlikely to be unambiguously resolved for some time, requiring both more integrated luminosity for the experiments and improved understanding of long-distance effects in the theoretical predictions.
This story is also seen in the tests of LFU in measurements of $R(D^{*})$, the ratio of branching fractions of $\Bz \to \Dstp\taum\antinutau$ and $\Bz \to \Dstp\mum\antinumu$. Such measurements are sensitive to new particles that couple preferentially to the heavier third generation of leptons, such as Higgs-like charged scalars or lepto-quarks.
With recent contributions from Belle~\cite{Sato:2016svk}, again with a unique contribution (reconstructing the $\tau$ decay hadronically rather than leptonically), the current world average, presented in Fig.~\ref{fig:rdrds}, is around $4\,\sigma$ away from the SM prediction~\cite{Amhis:2016xyh}.

\begin{figure}[!h!tbp]
  \includegraphics[width=0.47\textwidth]{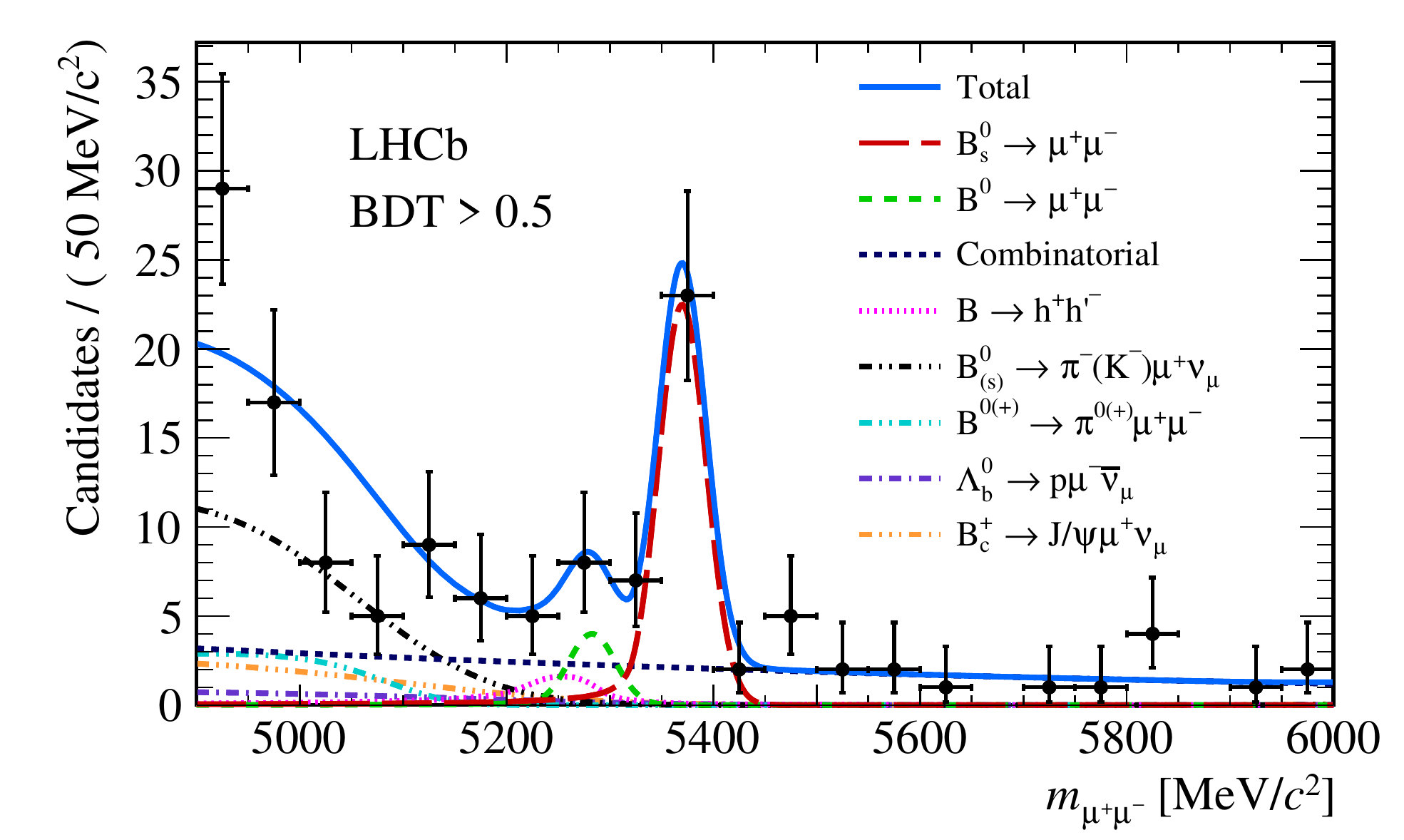}
  \hfill
  \includegraphics[width=0.43\textwidth]{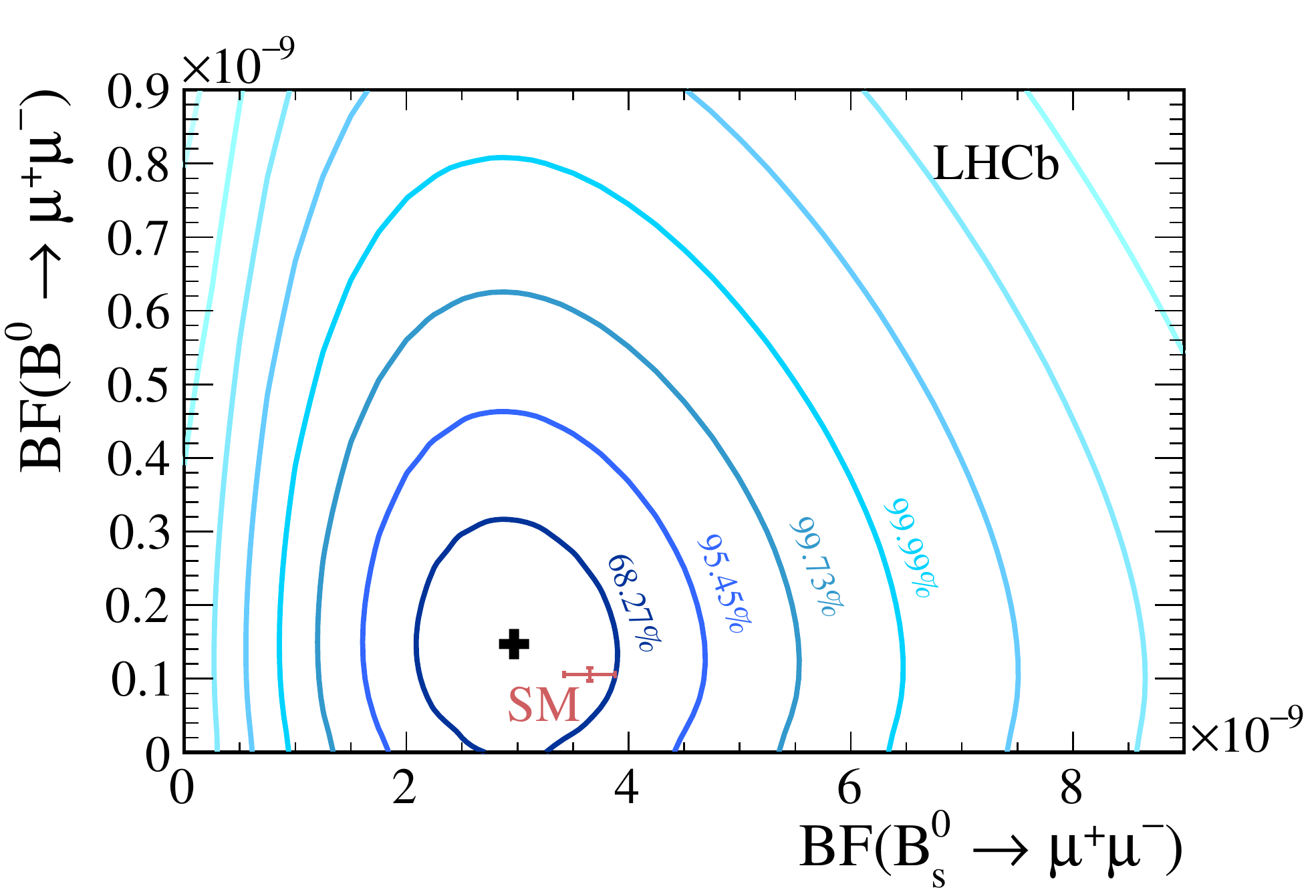}
  \caption{%
    (Left) Dimuon invariant mass distribution from the search for the $B_{(s)}^{0} \to \mup\mum$ decay. (Right) Comparison of the $\Bs \to \mup\mum$ measurement (red point) and $\Bz \to \mup\mum$ upper limits (in confidence intervals, as contours) with the SM prediction (black point)~\cite{Aaij:2017vad}.
  }
  \label{fig:bstomumu}
\end{figure}

Still, as a testament to just how much data has already been collected at the LHC, LHCb has made the first observation by a single experiment of the FCNC process $\Bs \to \mup\mum$~\cite{Aaij:2017vad}, shown in Fig.~\ref{fig:bstomumu}.
They report a branching fraction for this process of $3.0 \pm 0.6^{+0.3}_{-0.2}\times10^{-9}$, the rarest observed heavy flavour branching fraction to date, which is in agreement with expectations from the SM.

\section{Exotic states}
\label{sec:exotica}

\begin{figure}[!h!tbp]
  \includegraphics[width=0.5\textwidth]{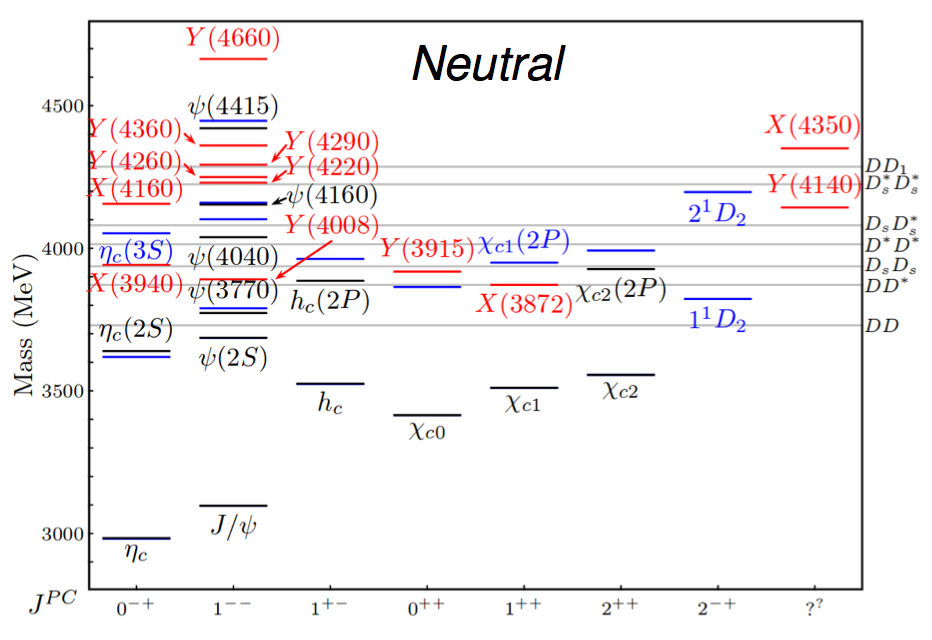}
  \hfill
  \includegraphics[width=0.5\textwidth]{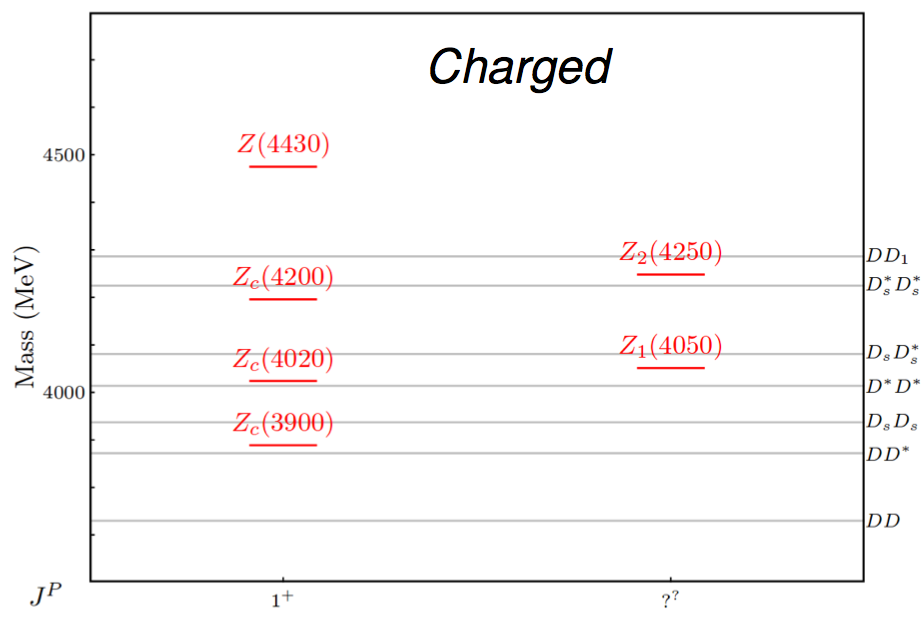}
  \caption{%
    Spectrum of most of the known exotic neutral (left) and charged (right) 
    states containing charm quarks. Notable omissions from the charged spectrum 
    are the two charm pentaquark states discovered by the LHCb experiment in 
    2015~\cite{Aaij:2015tga}.
  }
  \label{fig:exotics}
\end{figure}

The so-called exotic states are hadrons containing at least 4 quarks.
Although the existence of such particles was theorised by Gell-Mann in 
1964~\cite{GellMann:1964nj}, the exact nature of many of the multitude of 
discovered states, partially enumerated in Fig.~\ref{fig:exotics}, is still not well 
understood.
Possible explanations for any given state can include the `usual' bound states 
of quarks, molecules composed of multiple 2- or 3-quark components, or 
kinematic effects such as rescattering.

\begin{figure}[!h!tbp]
  \includegraphics[width=0.5\textwidth]{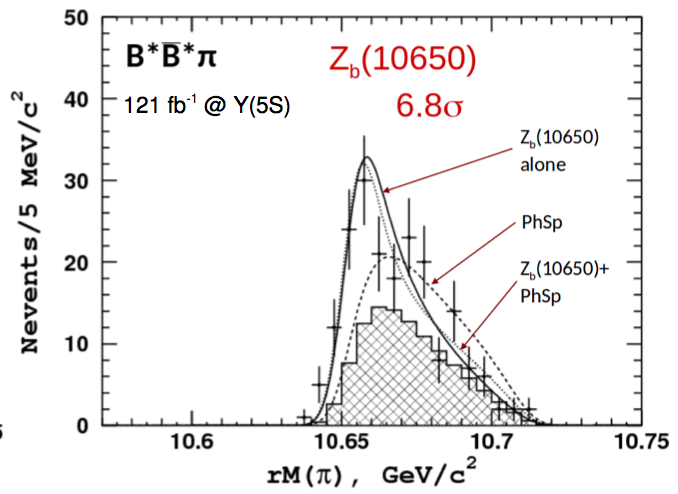}
  \hfill
  \includegraphics[width=0.5\textwidth]{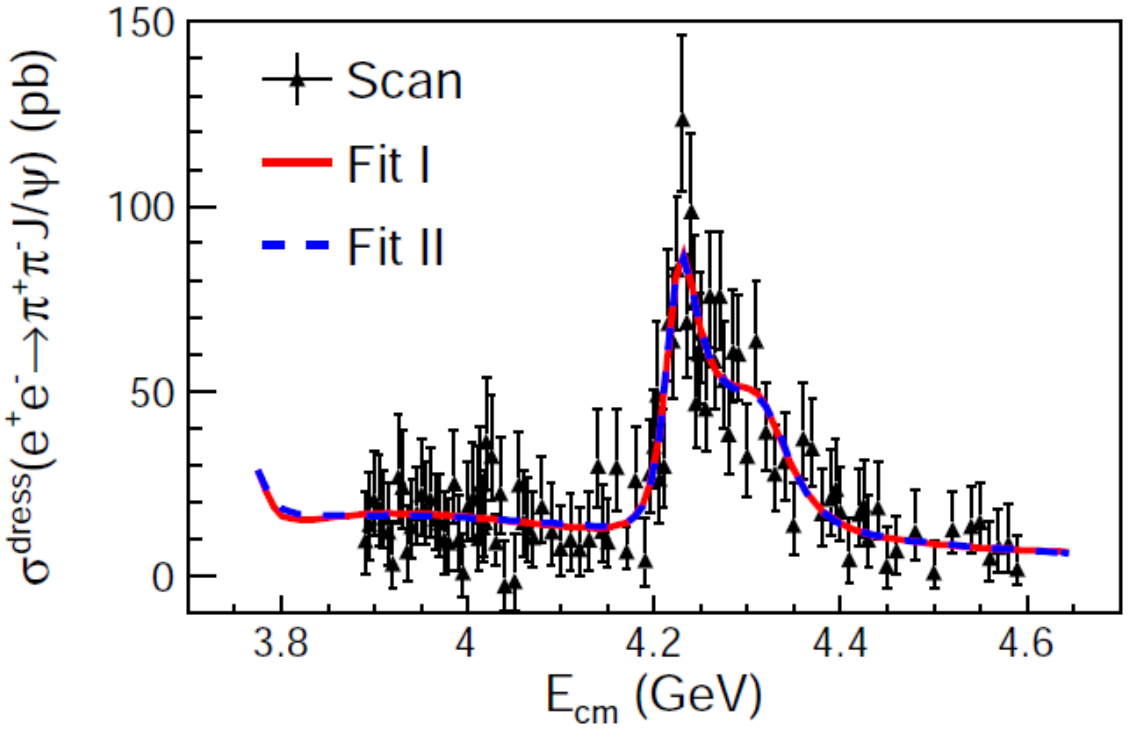}
  \caption{%
    (Left) Distribution of $M_{\mathrm{miss}}(\pi)$ for the $\electronp\electronm \to B^{*}\bar{B}^{*}\pi$ process~\cite{Garmash:2015rfd}, illustrating that the dominate component is through a resonant exotic state, $\electronp\electronm \to Z(10650)\pi$.
    (Right) Cross-section measurements for $\electronp\electronm \to \Jpsi\electronp\electronm$ made at a range of centre-of-mass energies, demonstrating two resolvable structures, reported to be the $Y(4260)$ and the $Y(4360)$~\cite{Ablikim:2016qzw}.
  }
  \label{fig:exotics-belle-bes}
\end{figure}

As well as studying the decay amplitudes of exotic states, experiments at 
$\electronp\electronm$ colliders can additionally probe their nature by 
performing scans across a range of centre-of-mass energies around some known 
invariant mass. Both \belle and \besthree presented such measurements~\cite{Garmash:2015rfd,Ablikim:2016qzw}, shown in 
Fig.~\ref{fig:exotics-belle-bes}, and expressed the need to now begin 
systematically characterising exotic states in order to understand their 
nature.

The need for futher theoretical understanding is highlighted by the recent discovery of the $X(5568)$ particle reported by the \dzero collaboration in the $\Bs\pipm$ final state~\cite{D0:2016mwd}. This was first performed using $\Bs \to \Jpsi\phi$, and a new preliminary analysis was presented at DIS'17 using semileptonic $\Bs \to \Dsp\mum X$ decays~\cite{bertram_dis17}. A local significance of $3.2\,\sigma$ was reported, which when combined with the initial measurement rises to $5.7\,\sigma$.
Intriguinly, neither LHCb nor CMS see the new state in their data~\cite{Aaij:2016iev,CMS-PAS-BPH-16-002}, despite \dzero finding a rate of $\Bs$ mesons produced via the $X(5568)$ of around $8\,\%$ in $p\bar{p}$ collisions. The \Bs samples produced by LHC are considerably larger than those of the Tevatron, and although it has been suggested that different collision environments and production mechanisms may account for at least some of the difference, the tension cannot be fully explained by existing theory.

\begin{figure}[!h!tbp]
  \includegraphics[width=0.5\textwidth]{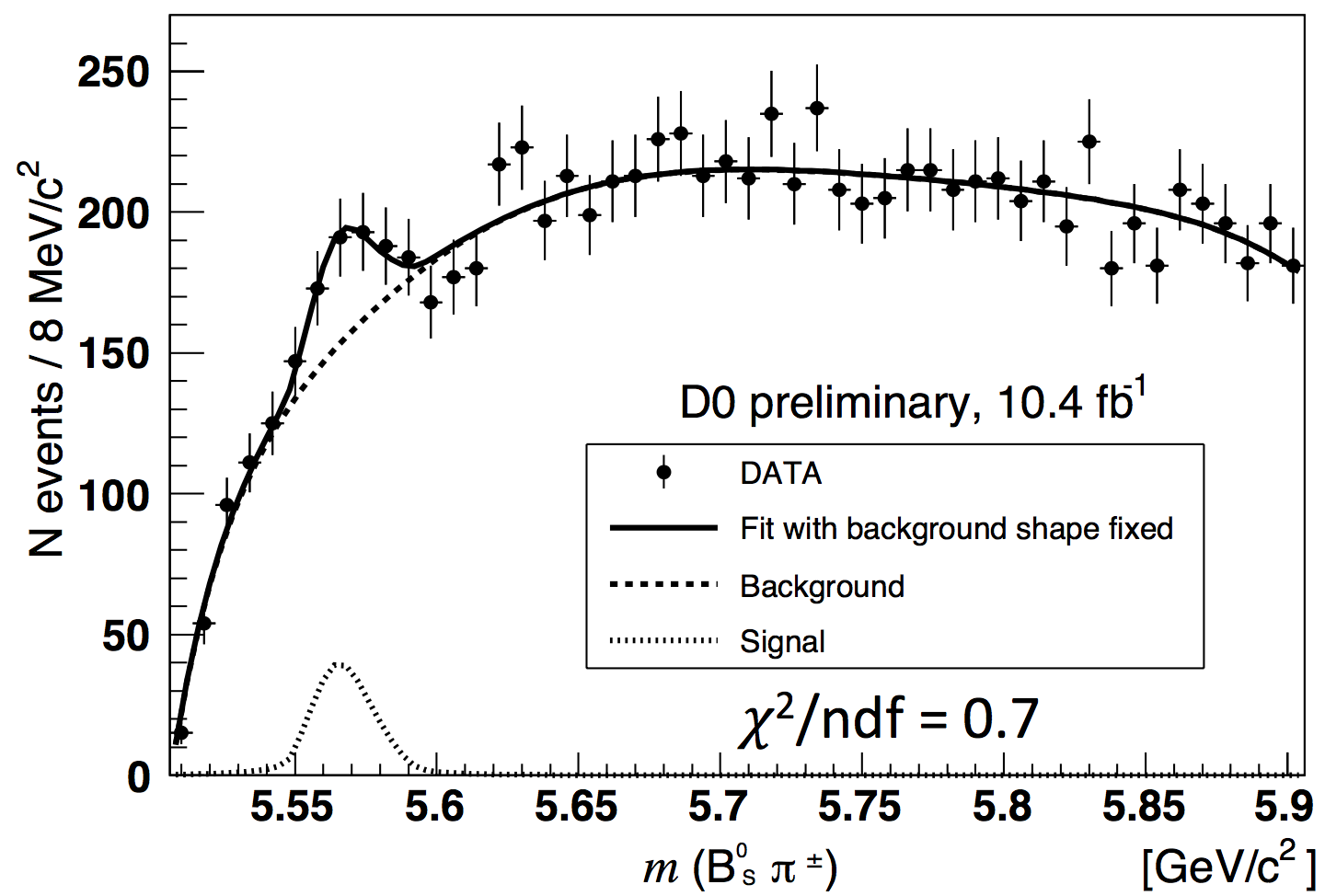}
  \hfill
  \includegraphics[width=0.5\textwidth]{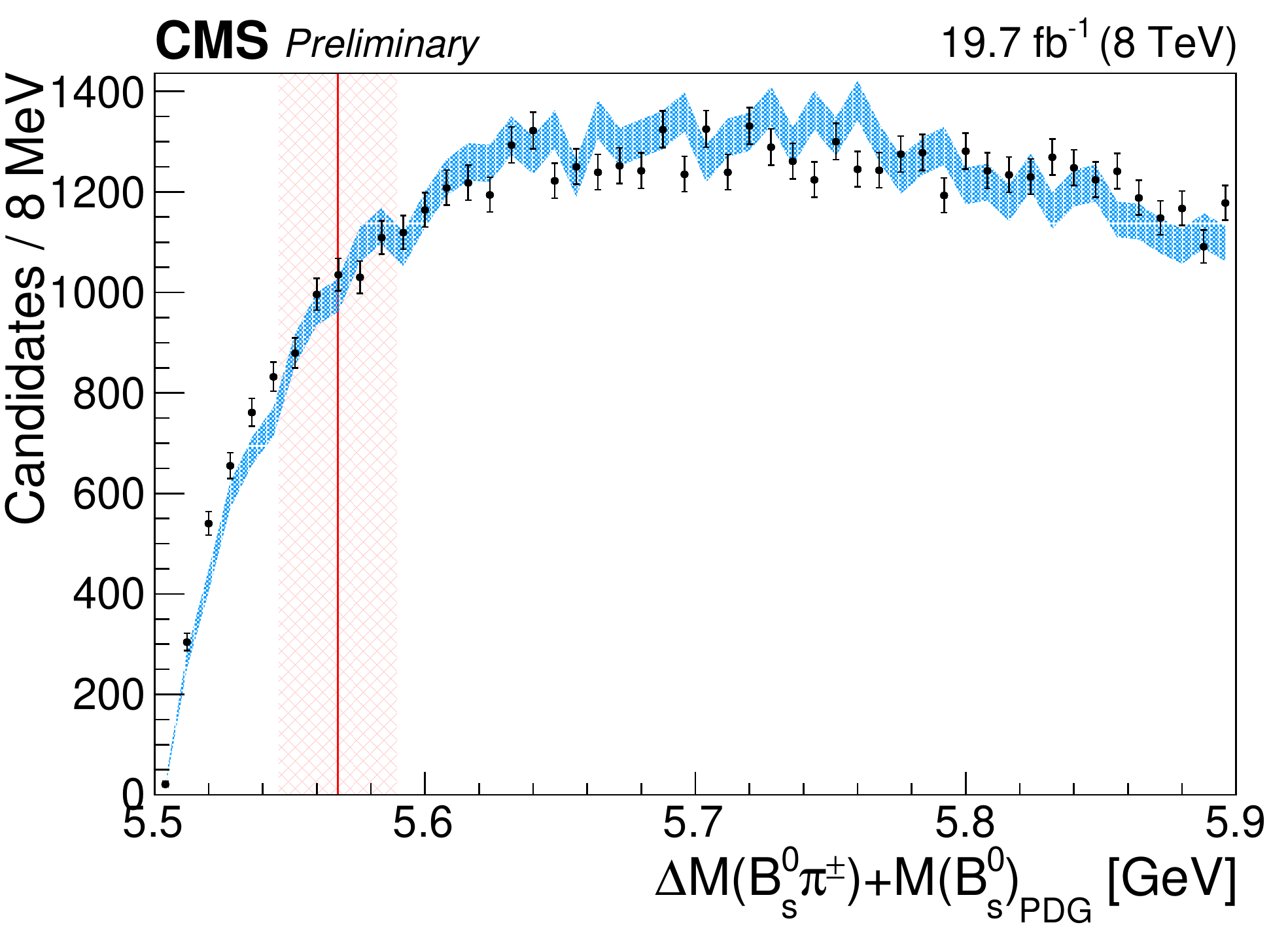}
  \caption{%
    (Left) Mass distribution $m(\Bs\pipm)$ as measured by the \dzero collaboration using the $\Bs \to \Dsp\mum X$ decay~\cite{bertram_dis17}. (Right) Corresponding mass range in the same final state in CMS data~\cite{CMS-PAS-BPH-16-002}.
    }
  \label{fig:x5568}
\end{figure}

\section{Quarkonia}
\label{sec:quarkonia}

Studies of quarkonia can reveal information both on the partonic structure of the colliding hadrons and on the evolution from the partonic initial state to the $q\bar{q}$ bound state. For \Jpsi production in particular, experimental data can help both disentangle and constrain the non-pertubative colour singlet and colour octet models that describe the hadronisation process.

\begin{figure}[!h!tbp]
  \includegraphics[width=0.5\textwidth]{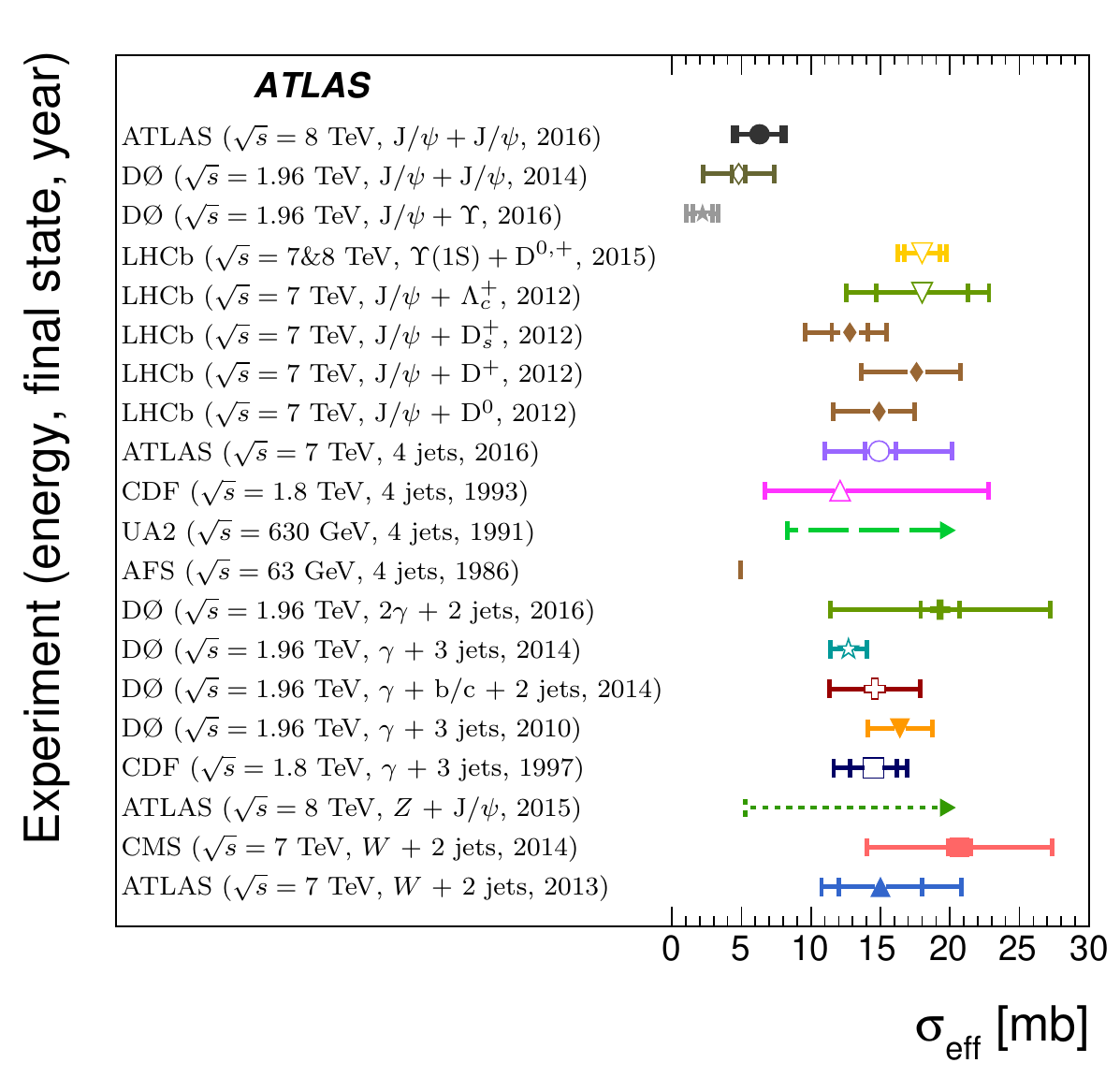}
  \hfill
  \includegraphics[width=0.5\textwidth]{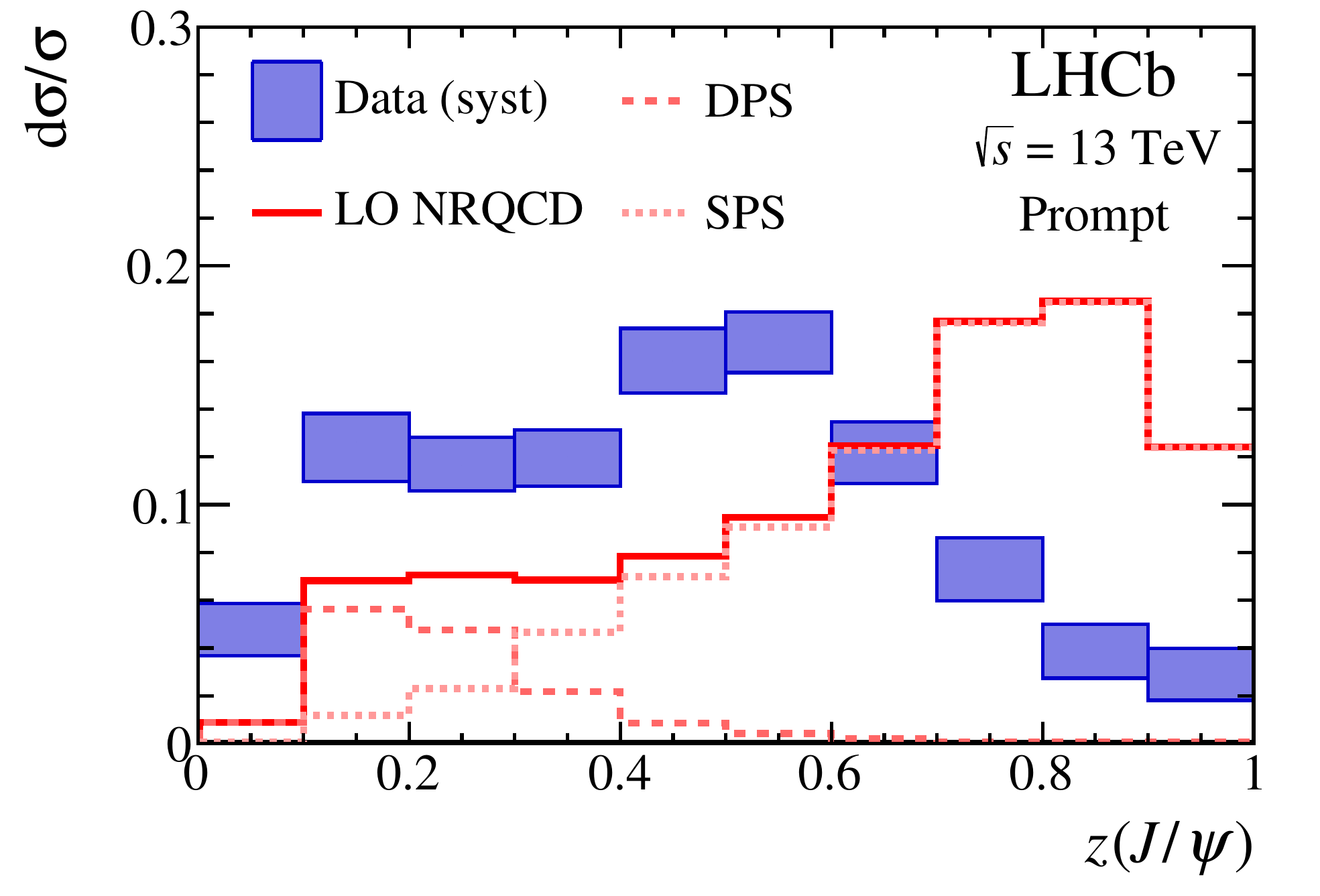}
  \caption{%
    Various experimental measurements of the effective double parton scattering cross-section, which is expected to be independent of collision energy and final state~\cite{Aaboud:2016fzt}.
    (Right) Relative \Jpsi production cross sections as a function of jet \pT that is carried by a \Jpsi internal to the jet~\cite{Aaij:2017fak}.
  }
  \label{fig:dps}
\end{figure}

The simultaneous production of two \Jpsi mesons can provide information on the process of double parton scattering~(DPS), where two independent hard processes occur in the same collision. In di-\Jpsi production, the contribution from leading-order computations for single parton scattering is far too small to account for the integrated cross-sections, and cannot describe the differential shapes in transverse momentum. The sum of single parton scattering (SPS) at next-to-leading order and DPS contributions \emph{can} describe the data~\cite{lansberg_dis17}, however, as demonstrated by a recent result from ATLAS~\cite{Aaboud:2016fzt}. A summary of recent double production measurements is given in Fig.~\ref{fig:dps} (left), which highlights the interesting result that the fraction of DPS to double SPS production is observed to be constant as a function of centre-of-mass energy and of the final state objects.

An additional test of QCD predictions can also be performed by measuring the production of \Jpsi within a reconstructed jet, and studying the relative transverse momentum fraction carried by the \Jpsi according to $z = \pT(\Jpsi)/\pT(\mathrm{jet})$.
Such a test can distinguish between non-relativistic QCD~(NRQCD) predictions, which usually predict an isolated \Jpsi, and those that model \Jpsi production in jets, either analytically or by using parton showers from event generators. This is particularly interesting because whilst differential \Jpsi production spectra are well-modelled by NRQCD, it also predicts a large transverse \Jpsi polarization, which has been not observed experimentally. The LHCb collaboration presented a meausrement of \Jpsi production in jets, measuring the momentum fraction $z$ for both \Jpsi mesons produced in $b$-hadron decays and those produced directly in the $pp$ collision~\cite{Aaij:2017fak}. Although the \Jpsi-from-$b$ momentum fraction distribution agree well with predictions, that for prompt \Jpsi production disgrees dramatically, as shown in Fig.~\ref{fig:dps} (right). As the first measuremnt of its type, it is a valuable contribution to understanding \Jpsi production in hadron collisions.

\section{Conclusion}

Working Group 5 of the DIS'17 conference hosted a wide variety of very interesting topics, and allowed for significant cross-pollination between usually dis-joint research areas.
  
At the LHC, the high-\pT general purpose experiments ATLAS and CMS, as well as the more specialised ALICE and LHCb, all show a larger degree of overlap as time progresses, with other colliders providing vital input in complimentary kinematic regions. 
With respect to few years ago, theorists interested in traditional DIS topics take inputs from many more sources, and experimentalists tackle the task of providing measurements with the highest utility.

The interplay of LHC heavy quark measurements with deep inelastic scattering is leading to greatly improved PDF precision, in turn allowing for, and requiring, better theoretical understanding. The  influx of heavy flavour data at the LHC is pushing down limits of rare decays, allowing for discoveries of new final states and exotic states, and opening up intriguing discrepencies in the weak sector. Experiments at other colliders are also producing superb results, very often complementary to those provided by the LHC experiments.

Although we have tried to cover as much as possible, the brevity of these proceedings requires us to omit many interesting topics that were presented. We encourage to reading to explore the other entries that discuss contributions made in the Physics with Heavy Flavour working group.

\section*{Acknowledgments}

We thank all the speakers and participants of Working Group 5 for very lively and insightful discussions, and the organisers of DIS'2017 for a great conference.

\bibliographystyle{JHEP}
\bibliography{wg5_proceedings}

\end{document}